\documentclass[conference]{IEEEtran}

\usepackage{graphicx}
\usepackage[colorinlistoftodos]{todonotes}

\usepackage{url}

\usepackage{hyperref}
\hypersetup{
    colorlinks=true,
    linkcolor=blue,
    filecolor=magenta,      
    urlcolor=blue,
}

\title{Cross-Platform Games in Kotlin}

\author{Simon M. Lucas \\ Game AI Research Group \\ 
School of Electronic Engineering and Computer Science \\
Queen Mary University of London}

\begin{document}

\thispagestyle{empty}
\pagestyle{empty}

\IEEEoverridecommandlockouts
\IEEEpubid{\begin{minipage}{\textwidth}\ \\[12pt] 978-1-7281-4533-4/20/\$31.00 \copyright 2020 IEEE \end{minipage}}

\maketitle

\begin{abstract}

This demo paper describes a simple and practical 
approach to writing cross-platform casual games 
using the Kotlin programming language.  A key
aim is to make it much easier for researchers to demonstrate their AI playing a range of games.
Pure Kotlin code (which excludes using any
Java graphics libraries) can be transpiled 
to JavaScript and run in a web browser.
However, writing Kotlin code that will
run without modification both in a web browser
and on the JVM is not trivial; it requires strict
adherence to an appropriate methodology.  
The contribution of this paper is to provide
such a method including a software design
and to demonstrate this working for Tetris,
played either by AI or human.

\end{abstract}

\section{Introduction}

The main motivation for this work is to enable 
researchers in games and game AI to publish
demonstrations of their games and algorithms that will run in a web browser or on a Java Virtual Machine (JVM).
The aim is to make the process as simple as possible
with minimal overhead.  
Running in a web browser enables interested viewers
to run the games and algorithms without the often
significant obstacle of having to install software.

When running in a web browser is not required,
the software can be run on a JVM.  JVM operation
is often around $10$ times faster than running in
a web browser using JavaScript.  JavaScript (and graphics using HTML5 Canvas)
is the only language that has widespread support
across all main web browsers without requiring
plugins (as would be necessary for Unity games for example).  Java Applets used to be an
alternative but running these is no longer
feasible as most browsers have disabled their support 
by default, and only allow them to be run with
the provision of security certificates.

The approach we take is to propose a software
design to enable the main game application
to be written in a platform independent way,
and provide an implementation of the approach 
in the Kotlin programming language.  A set
of classic 2D games will be provided as samples,
all with fast and easily copied
forward models, which are ready for statistical
forward planning algorithms such as Monte Carlo Tree
Search and Rolling Horizon Evolution.

Kotlin is chosen for the following reasons:
\begin{itemize}
\item Modern high-level language with many features that support good programming practice.
\item Fast execution (comparable to Java or C++).
\item Transpiles to JavaScript.
\item Recommended language for programming on the Android platform.
\item Support for Domain Specific Languages (DSLs), well suited for easy and type-safe implementation of game description languages.
\end{itemize}

Using Kotlin does not in itself lead to programs that will run both on the JVM and in a web browser.
Reasons include:
\begin{itemize}
    \item The graphics libraries of Java and of JavaScript are different, although they offer similar powerful features.  
    \item The event handling is similar, but differs in the details.
    \item Many of the Java libraries (such as Collections) have pure Kotlin equivalents, but they are not all the same, so care must be taken to only use pure Kotlin.  Currently the pure Kotlin random number generator offers similar features to the Java one, but is missing a \verb+nextGaussian()+ method for example.
    \item Multi-threading is harder in JavaScript.
\end{itemize}

As a solution, this paper proposes XKG: Cross (X) platform Kotlin Games.  XKG comprises a set of interfaces and classes, together with a set of guidelines
and examples of how to write cross-platform games.

\section{XKG: Cross-platform Kotlin Games}

\begin{figure}[hbtp]
\centering
\includegraphics[width=1.0\linewidth]{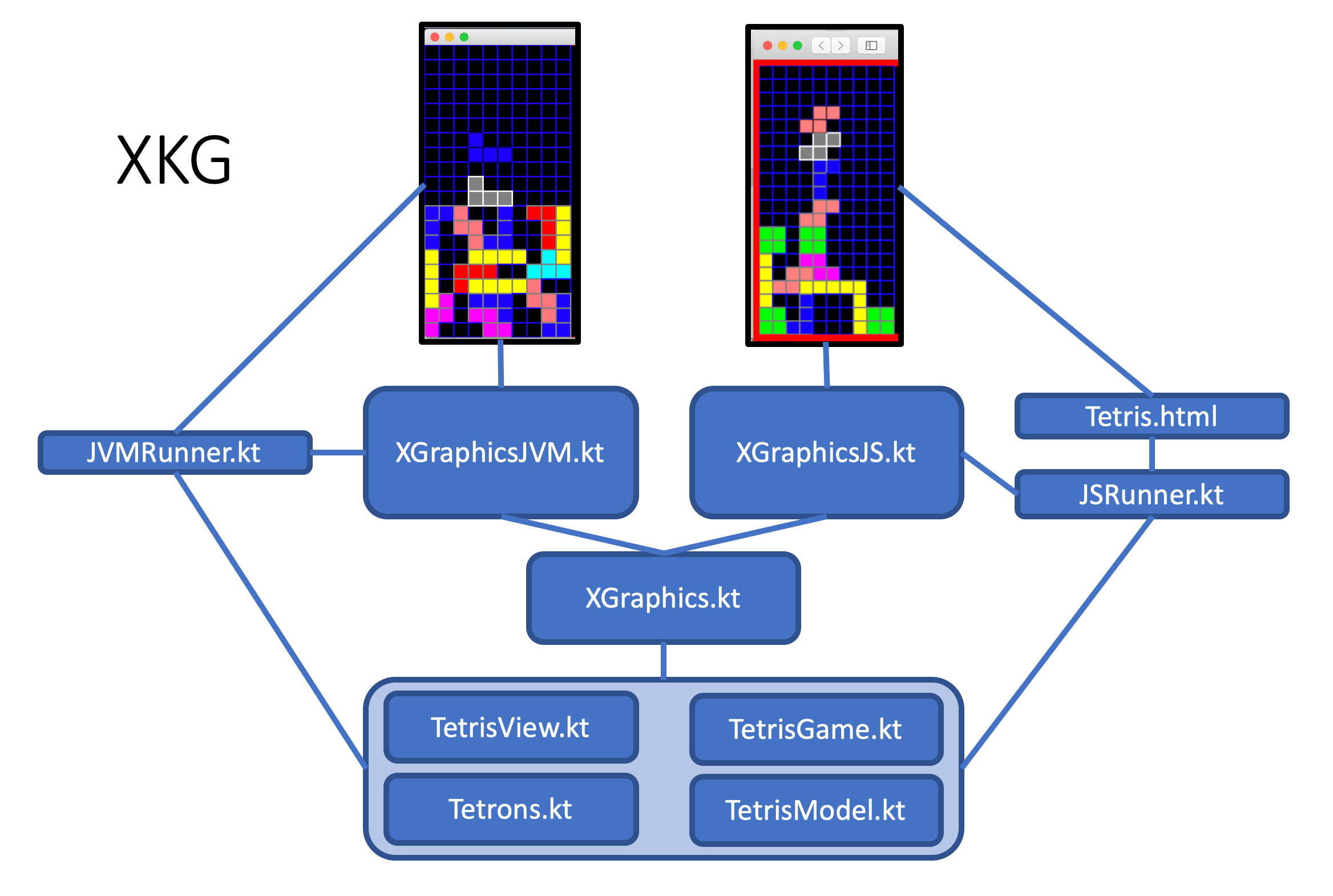}
\caption{\label{fig:XKG}
System diagram shown with an example implementation
of Tetris.  XKG enables the bulk of
the game or application code to be entirely platform independent.  In the case of Tetris, all the code specific to the game is shown in the light blue rectangle (TetrisView.kt etc.).  The rest of the code remains the same for any game.
Each game has its own simple HTML file (here Tetris.html) which launches the selected game.}
\end{figure}

The architecture of an XKG game is shown
in figure~\ref{fig:XKG}, and the main
modules are described below, with lines of source code of each in parentheses.  All code is written in Kotlin. The only non-Kotlin file is the HTML launcher page.

\textbf{XGraphics.kt (112)} Set of platform independent interfaces, enumerations and data bound classes.  All graphics used in cross-platform games go through XGraphics.  The data bound classes describe geometric
shapes (Lines, Polygons, Text etc.) and rendering options, and the interfaces provide methods for drawing these.

\textbf{XGraphicsJVM.kt (171)} Implementation classes of XGraphics interfaces that run on the JVM.

\textbf{XGraphicsJS.kt (139)} As above, but run in a Web Browser.

\textbf{JVMRunner.kt (24)} Small program to create a window (a JFrame), create the specified game (which could be passed as a command-line argument) and hook up the event handling, and start the main game loop (which is always game independent).

\textbf{JSRunner.kt (242)} Similar to the above, except to run in the browser.  For convenience, the HTML file specifies the game to run, and the JSRunner extracts
that from the HTML file and launches the specified game.
The lines of code for this is quoted as is, and
includes temporary experimental code providing bespoke 
launching of other test games.

\subsection{Demo}

The project is available on Github\footnote{\url{https://github.com/SimonLucas/XKG}}  enabling anyone to run the JVM versions as well as the Web versions of the games, and also to use and extend
as needed under a GPL-3.0 License.
In addition to Tetris (figure~\ref{fig:XKG}), 
there are also versions of Asteroids and
the word game Griddle.\footnote{\url{https://play.google.com/store/apps/details?id=com.lucapps.gstate&hl=en}}  
All games are
playable by human or AI.  The demo includes
visualisations of the games being played with
Rolling Horizon Evolution \cite{RalucaAnalysisRHEA} (showing the expected reward
for each rollout over time).  More visualisations
are currently being prepared.  One of the aims of XKG
is to provide an easy way to publish informative
interactive articles in the style of Distill\footnote{\url{http://distill.pub}}, 
but with support tailored for games and game AI.

\subsection{Timing Results}

Table~\ref{tab:speed} shows the number
in millions of game ticks (state transitions) per second achievable
using this platform when run
headless.  
\footnote{Macbook Pro 2.3 GHz Intel Core i5.}

Heuristic Tetris incorporates a column height difference penalty in the score which makes the game much easier for reinforcement learning and statistical forward planning algorithms.  The current calculation of column difference is done inefficiently on every call to the score function, and is only included here to provide 
another data point for timing comparison.

The transpiled JavaScript
running in the web browser is around 8 times slower
than the code running on the JVM, but still
achieves around 1.4m ticks per second,
which is more than what's needed for typical 
SFP algorithms.  Of course, faster is always better
for running experiments, which is why we recommend
running on the JVM for generating experimental
results, and then running in a web browser for
easy demonstrations and play-testing of games.

\begin{table} [!t]
\centering
\caption{\label{tab:speed}
Millions of game ticks per second achieved for
some sample games that come with the platform.
Tetris is the standard game where points are only awarded for each line completed.  Heuristic Tetris is explained in the text.}

\begin{tabular}{lrr}
\hline
Game &  JVM & JavaScript (4) \\
\hline
Tetris & 10.0 & 1.3 \\
Heuristic Tetris & 1.0 & 0.2 \\
\hline
\end{tabular}
\end{table}



\section{Discussion and Summary}

We are currently in the process of adding a wider
selection of games to the platform and extending the
range of graphics and other features on a per-need basis.  So far the experience of doing this has
raised a few frustrations but for the most part
has been straightforward and enjoyable.  
Surprising difficulties
included simple things, such as colours rendering differently in the browser compared to the JVM.
A feature of web browsers is that they may apply
their own colour space model so that specifying
a colour as pure red in RGB coordinates can
translate to an impure red, a fact easily verified
(e.g. on a Mac, the Digital Color Meter utility can be used).


Future developments including enhancing the platform's
core features by adding support for audio and network
connections, and extending the range of games available.
Adding support for a general game platform \cite{GVGAISurvey} \cite{LudiiComparison}
would vastly
increase the number of games available, and by using
Kotlin's DSL support it may be possible to achieve this
without loss of speed.  Two-player games such
as a tunable version of Planet Wars \cite{LucasPlanetWars}
where where players can compete with the AI
will also be added.  Also to be added is a demo
of sample efficient optimisation (of agents or games), starting with NTBEA \cite{lucas2018n}.

Providing interactive demonstrations and games
that people can play with minimal effort can
help with the rapid proliferation of new Game AI ideas
and is clearly desirable.  This paper
introduced an approach to enable this using Kotlin, a powerful modern programming language.
As explained in the paper, while Kotlin provides
a great starting point for writing cross-platform
games, it is not enough by itself.  XKG provides
a simple and practical solution.

\bibliographystyle{IEEEtran}

{\small \bibliography{spindemo}}

\begin{thebibliography}{1}
\providecommand{\url}[1]{#1}
\csname url@samestyle\endcsname
\providecommand{\newblock}{\relax}
\providecommand{\bibinfo}[2]{#2}
\providecommand{\BIBentrySTDinterwordspacing}{\spaceskip=0pt\relax}
\providecommand{\BIBentryALTinterwordstretchfactor}{4}
\providecommand{\BIBentryALTinterwordspacing}{\spaceskip=\fontdimen2\font plus
\BIBentryALTinterwordstretchfactor\fontdimen3\font minus
  \fontdimen4\font\relax}
\providecommand{\BIBforeignlanguage}[2]{{%
\expandafter\ifx\csname l@#1\endcsname\relax
\typeout{** WARNING: IEEEtran.bst: No hyphenation pattern has been}%
\typeout{** loaded for the language `#1'. Using the pattern for}%
\typeout{** the default language instead.}%
\else
\language=\csname l@#1\endcsname
\fi
#2}}
\providecommand{\BIBdecl}{\relax}
\BIBdecl

\bibitem{RalucaAnalysisRHEA}
R.~D. Gaina, J.~Liu, S.~M. Lucas, and D.~Pérez-Liébana, ``{Analysis of
  vanilla rolling horizon evolution parameters in general video game
  playing},'' \emph{European Conference on the Applications of Evolutionary
  Computation}, pp. 418 -- 434, 2017.

\bibitem{GVGAISurvey}
D.~Perez-Liebana, J.~Liu, A.~Khalifa, R.~D. Gaina, J.~Togelius, and S.~M.
  Lucas, ``{General Video Game AI: a Multi-Track Framework for Evaluating
  Agents, Games and Content Generation Algorithms},'' \emph{IEEE Transactions
  on Games}, vol.~11, pp. 195--214, 2019.

\bibitem{LudiiComparison}
E.~Piette, M.~Stephenson, D.~Soemers, and C.~Browne, ``{An Empirical Evaluation
  of Two General Game Systems},'' \emph{IEEE Conference on Games}, 2019.

\bibitem{LucasPlanetWars}
S.~M. Lucas, ``{Game AI Research with Fast Planet Wars Variants},'' \emph{IEEE
  Conference on Computational Intelligence and Games}, 2018.

\bibitem{lucas2018n}
S.~M. Lucas, J.~Liu, and D.~Perez-Liebana, ``{The N-Tuple Bandit Evolutionary
  Algorithm for Game Agent Optimisation},'' \emph{IEEE Congress on Evolutionary
  Computation}, 2018.

\end{thebibliography}

\end{document}